\documentstyle[amssymb,aps,prl,multicol,epsf]{revtex}

\begin{document}

\title{Universal transport in 2D granular
superconductors}
\author{$^{1}$A.Frydman $^{2}$O. Naaman and $^{2}$R.C. Dynes}
\address{$^{1}$Department of physics, Bar Ilan university, Ramat Gan, 52900, Israel \\
$^{2}$The Department of Physics, University of California, San Diego, La
Jolla, CA 92093.}
\date{\today}
\maketitle

\begin{abstract}
The transport properties of quench condensed granular superconductors are
presented and analyzed. These systems exhibit transitions from insulating to
superconducting behavior as a function of inter-grain spacing.
Superconductivity is characterized by broad transitions in which the
resistance drops exponentially with reducing temperature. The slope of the
log \newline
R versus T curves turns out to be universaly dependent on the normal state
film resistance for all measured granular systems. It does not depend on the
material, critical temperature, geometry, or experimental set-up. We discuss
possible physical scenarios to explain these findings.


\end{abstract}

\begin{multicols}{2}
Two dimensional granular superconductors, i.e. systems of superconducting
grains embedded in an insulating matrix, exhibit a superconductor to
insulator transition (SIT) as a function of the inter-grain distance. Such
systems are of interest since they can serve as a model system for many
dirty superconductors. In particular it has been suggested that a granular
superconductor can mimic the behavior of high T$_{C}$ superconductors and
its properties can be used to explain some of the features observed in the
cuprates \cite{lynne}.

An established technique to study the properties of granular superconductors
is quench condensation \cite{strongin,granular bob,granular goldman,granular
rich,valles stm}. In this method one performs sequential evaporations on a
cryogenically cold substrate under UHV conditions using the following
scheme: Metallic leads for four-terminal measurements are prepared on an
insulating substrate which is then mounted onto an evacuated measurement
probe and immersed in a liquid He bath. The low temperature of the probe
causes cryopumping and hence the substrate is situated in UHV conditions
and, at the same time, at cryogenic temperatures. This enables evaporation
of ultra-clean thin superconducting layers on a substrate held at
temperatures lower than 10K while continuously monitoring the film
resistance and thickness. Once a desired resistance (or film thickness) is
achieved, the evaporation is terminated and the transport properties are
measured. Incremental layers of material are then added {\em in-situ} and
further measurements are taken at different film resistances. Using this
method one can study the properties of a single sample throughout the entire
transition from an insulator to a superconductor as a function of the amount
of deposited material while keeping the sample at low temperatures and in a
UHV environment without having to thermally cycle it (risking metallurgical
or structural changes due to annealing) or to expose it to atmosphere (thus
oxidizing the surface).

If the samples are quench condensed on a passivated substrate (such as SiO$%
_{2}$), they grow in a granular manner so that the film begins its growth as
disconnected islands with diameters of 100-200 \AA\ \cite{valles
stm,tony,philmag}. The average distance between the islands decreases upon
adding material. Beyond a percolation threshold, the grains connect, forming
a continuous conducting layer. In these samples there is a critical nominal
thickness, $d_{C}$, below which no conductivity can be measured (the sheet
resistance, R, is larger than 10$^{10}\Omega )$. Once the
thickness, d, of the sample is larger than $d_{C},$ R drops
exponentially with thickness until, for $R \leq 6k\Omega ,$%
\ it crosses over to a normal ohmic behavior ($R \alpha 1/d$) %
\cite{philmag}.

Varying the thickness of the film causes a transition from an insulating
behavior for the thinnest films to a superconducting behavior for thick
films. Figure 1 demonstrates examples for this transition in three different
granular superconductors: a Pb film (critical temperature, $T_{C},\approx $%
7.2 K), a Sn film ($T_{C}\approx $4.5K) and a Pb/Ag bilayer. The latter is a
system in which a thin layer of insulating granular Pb is quench condensed
on a SiO$_{2}$ substrate followed by sequential evaporations of Ag ultrathin
layers \cite{lynne,pss}. The curves were measured in a shielded room using
standard 4 probe technics and assuring, for each point on the curve, that
the I-V characteristics are in the linear regime.

The three systems show a transition from an insulator for thin films to a
superconductor at thicker films. In all 2D granular systems these
transitions are characterized by broad resistance tails and critical
temperatures which are not very well defined until the resistance of the
film becomes low. The transitions become sharper as material is added to the
film. In this paper we define $T_{C}$ as the temperature at which the
resistance starts dropping exponentially with lowering the temperature. We
justify this by observing that this is the temperature in which the
individual grains become superconducting even in the insulating case as
discussed below. In the Pb and Sn films (top two graphs in figure 1) $T_{C}$
has bulk value and barely changes throughout the entire transition.
Moreover, even on the insulating side of the SIT the curve changes its slope
at $T=T_{C}$ reflecting the presence of superconductivity even in the
thinnest measurable samples. The fact that the grains are superconducting
with bulk properties throughout the entire transition has been demonstrated
by tunneling measurements \cite{richlynne}. A bulk energy gap was observed
in the grains even when the film was on the insulating side of the SIT.
\begin{figure}[b]
\centerline{
\epsfxsize=100mm
\epsfbox{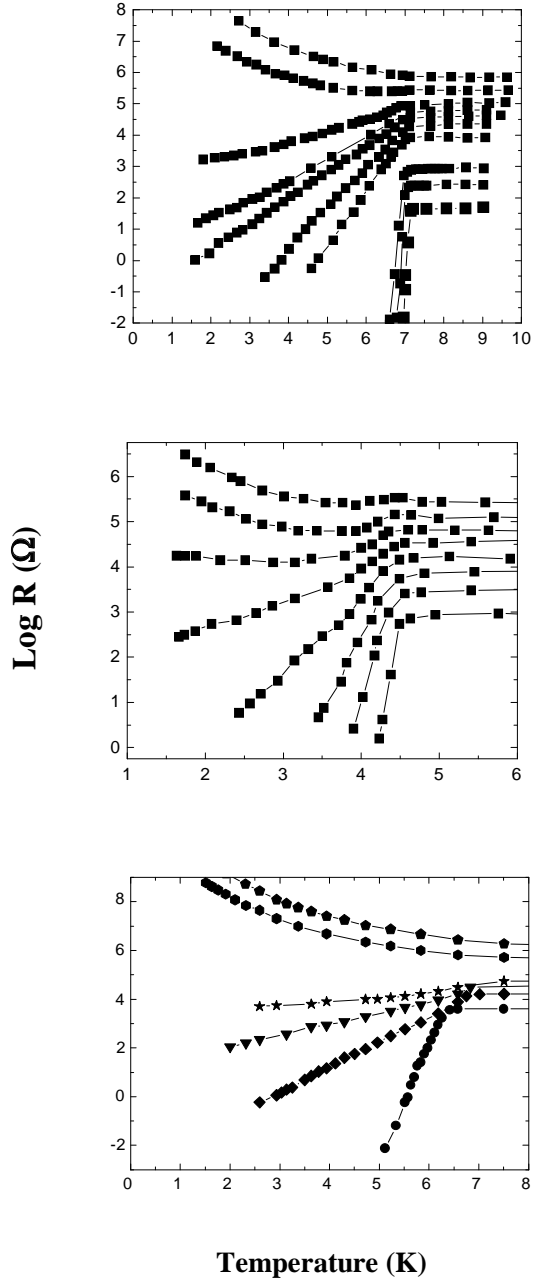}
}
\vspace{0mm}
\narrowtext
\caption{Resistance versus temperature for sequential layers of
quench-condensed granular Pb (Top), Sn (middle) and Pb/Ag (bottom).
Different curves correspond to different nominal thickness.
 }
\end{figure}

In the Pb/Ag film the $T_{C}$ decreases as material is added to the film.
This is due to suppression of superconductivity in the Pb grains by the
silver overlayer because of proximity effects. Nevertheless, the general
behavior of the insulator to superconductor transition is very similar to
that of the pure superconductors.

The transport curves shown in figure 1 are attributed to the unique nature
of the SIT in a granular structure \cite{compare}. In these systems each
grain sustains superconductivity with bulk properties. However, for the high
resistance samples, there are rapid, thermally activated phase fluctuations
between the grains leading to an insulating R-T curve. As the grains become
closer the inter-grain Josephson coupling increases until phase coherence is
achieved throughout the film.

A striking feature common to the quench condensed granular superconductors
is that at temperatures below $T_{C}$ the sheet resistance decreases
exponentially with decreasing temperature and can be described approximately
by an ''inverse arrennius'' law:

\begin{equation}
R=R_{0}e^{\frac{T}{T_{0}}}  \label{eq1}
\end{equation}

It is important to note that we see no flattening of the R-T curve
( resistance approaching a flat temperature dependence at low temperatures)
in any of our samples and in no regime of thicknesses \cite{ref goldman}.
Nor does the R-T change its trend and begin dropping to zero resistance at
low temperature. The dependence described in equation 1 is observed in all
of our granular samples using different materials and different measurement
apparatuses. It spans many orders of magnitude in R, persists to
temperatures below 100 mK \cite{richbob} and is observed for the different
steps of the sequential evaporations providing the resistance is smaller
than a few tens of $k\Omega $. Moreover, the slope of the log R versus T
curves, $1/T_{o}$ of equation 1, turns out to be universal for all of our
samples. It depends only on the normal state sheet resistance, $R_{N},$ and
does not depend on the material, the critical temperature or sample
geometry. We illustrate this in Figure 2 where we show the dependence of the
inverse slope, $T_{0},$ on $R_{N}$ for a large number of granular
superconductors (different materials, geometries and prepared in various
quench condensation evaporators). It is seen that all the slopes fall on a
master-plot having the form of:

\begin{equation}
T_{0}\approx C^{\ast }R_{N}  \label{eq2}
\end{equation}

where C* is a constant of approximately 0.05$\frac{K}{k\Omega }.$ Note that
we have included samples having different critical temperatures as well as
the Pb/Ag system in which $T_{C}$ varies with the thickness (and hence with $%
R_{N}$).

This observation indicates that the behavior of the superconducting tails
does not depend on the properties of the superconducting grains themselves
but only on their geometrical arrangement (density, inter-grain spacing,
morphology configuration etc.) which determines the tunneling percolation
network and $R_{N}$. \bigskip Two superconductors having the same $R_{N}$
but different critical temperatures $T_{C1}$ and $T_{C2}$ will have parallel
superconducting R-T tails, at temperatures lower than the respective $T_{C},$
shifted by $T_{C1}-T_{C2}.$ An example for such behavior is shown in the
insert of figure 2.
\vspace{0mm}
\begin{figure}[b]
\centerline{
\epsfxsize=150mm
\epsfbox{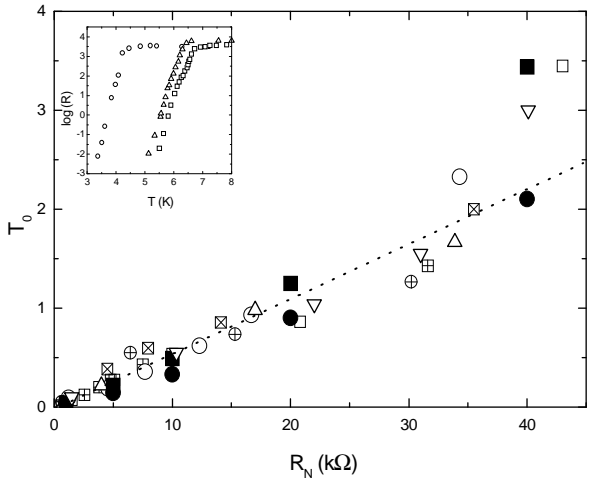}
}
\vspace{-10mm}
\narrowtext
\caption{The inverse slope of the R-T curves ( $T_{O}$ of equation 1)
as a function of the normal-state sheet resistance for different granular
samples. The open symbols are the experimental data. Squares are results for
various Pb samples, circles are for Sn samples and triangles for Pb/Ag
proximity samples. Single symbols are used for the same quench-condensed
films at different thicknesses (and hence R$_{N}$s). Full symbols are the
slopes extracted from the calculated samples of figure 3 (squares -Pb and
circles - Sn). The line is a guide to the eye.
Insert: Resistance versus temperature for granular Pb (squares), Sn
(circles) and Pb/Ag (triangles) samples having a normal state resistance of
4 k$\Omega .$
}
\end{figure}

We can now combine this observation with equations 1 and 2 and extract the
expression for R$_{0}$ of equation 1:

\begin{equation}
R_{0}=R_{N}e^{\frac{-T_{C}}{T_{0}}}=R_{N}e^{\frac{-T_{C}}{C^{\ast }R_{N}}}
\label{eq3}
\end{equation}

R$_{0}$ is the value of the resistance obtained by extrapolating the R-T
curves to T=0. This means that if the observed exponential R-T curves
persist to zero temperature the samples will exhibit a finite
zero-temperature resistance which will depend both on $R_{N}$ and on $T_{C}$.

In an attempt to examine possible intuitive understanding of the observed
phenomena we note that at temperatures below bulk $T_{C}$ the grains have
been shown to be fully superconducting \cite{richlynne}. Hence, each two
grains are expected to be Josephson coupled with a Josephson binding energy
of the form:

\begin{equation}
Ej=\frac{\pi \hbar }{4e^{2}}\frac{\Delta (T)}{R_{N}}\tanh \frac{\Delta (T)}{%
2K_{B}T}  \label{eq4}
\end{equation}

where $\Delta (T)$ is the temperature dependent superconducting gap and R$%
_{N}$ is the normal resistance between the grains. One can expect that the
ratio $\frac{Ej}{K_{B}T}$ would determine whether the grains are phase
coupled or not. As the temperature of a granular system is lowered Ej
becomes larger than K$_{B}$T in an increasing number of pairs of grains thus
increasing the effective superconducting regions which are phase coherent.
This leads to a characteristic ''phase-coupled'' length which grows with
decreasing temperature until superconducting percolation is achieved.

However, the naive model described above can not explain the observed broad
superconducting tails in our films. Since the samples are two dimensional,
scaling considerations imply that as long as superconducting percolation is
not achieved throughout the sample, the resistance of the granular system
should be independent on the size of the superconducting clusters. The
growth of the phase coherent clusters with decreasing temperature would not
change the overall 2D resistance. One can expect to see a flat resistance
versus temperature curve below the percolation threshold and a sharp
superconducting transition at the percolation threshold. The fact that the
experimental transitions are broad implies that the resistance of the
individual junctions is temperature dependence. This urges us to consider
superconductor fluctuation effects as fluctuations introduce a broad
temperature dependence in a Josephson junction resistance rather than a
sharp transition at Ej=K$_{B}$T.

The temperature dependence of the zero-bias resistance of each pair of
grains due to thermal phase fluctuations is expected to take the
Ambegaokar-Halperin form \cite{halperin,ofer}:

\begin{equation}
R(T)=\frac{R_{N}}{(I_{O}(\frac{Ej(T)}{K_{B}T}))^{2}}  \label{eq5}
\end{equation}

where I$_{0}$(X) is the modified bessel function of order 0. Using this
model we have calculated the temperature dependence of 1D arrays of
junctions subject to thermal fluctuations. The results of the simulations
for Pb and Sn samples having different $R_{N}$ are shown in figure 3. It is
seen that R(T) may qualitatively mimic an exponential dependence for
temperatures larger than about $\frac{T_{C}}{2}$. As $R_{N}$ is reduced the
R-T curves become sharper in a similar way to that seen in the experiments.
While the curves are not truly e$^{T}$ dependent, for purposes of
illustration we can extract an approximated slope from each curve. The
extracted slopes are plotted on the master-plot of figure 2 showing
relatively good agreement with the experimental results. We can not expect
better than qualitative agreement as this is a simple 1D chain model.

\begin{figure}[b]
\centerline{
\epsfxsize=90mm
\epsfbox{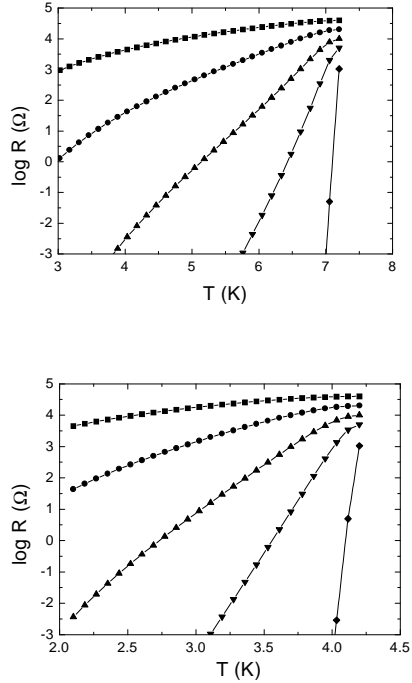}
}
\vspace{-0mm}
\narrowtext
\caption{Numeric simulations of the resistance of arrays of 10
junctions in series based on equation 5. The top graph is for Sn and the
bottom for Pb.
}
\end{figure}

Hence, the thermal-fluctuation-picture can provide some insight to the
physics of the 2D granular films. However, two major points in which the
simple model deviates from the experimental data have to be noted: In the
first place the theoretical model predicts that the slopes should depend on $%
\frac{T}{T_{C}}.$ Our simulations fail to reproduce the $T_{C}$ independence
and the ''universality'' of the experimental transition tails. The
calculated curves for Sn and Pb samples having the same $R_{N}$ yield Sn
slopes which are consistently sharper than those of Pb. Secondly, the
thermal-fluctuation model can qualitatively imitate the experimental
behavior only for a limited range of temperatures. At low temperatures the
thermal fluctuations are expected to freeze out causing the resistance to
drop much more rapidly, until, for zero temperature, the resistance reaches
zero ( $R_{0}$=0). We do not see signs for such a tendency in the
experiments even at very low temperatures. We have considered the
possibility that at low temperatures quantum phase fluctuations dominate the
behavior and induce a finite resistance even at T=0. The presence of such
effects is consistent with the observed experimental dependence of $R_{0}$
on $T_{C}$ as shown in equation 3 since the quantum fluctuations are
expected to decay exponentially with Ej. However, we are not able to
reproduce the experimental exponential R-T curves over the entire
temperature range using a model which combines thermal fluctuations and
quantum fluctuations. Such a model would suggest that at low temperature the
R-T curve would flatten out and saturate at a constant value. As noted
above, we do not observe such behavior in any of our samples. Clearly,
further theoretical treatment is required in order to shed more light on the
origin of the findings presented in this paper.

This research was supported by the Israel - USA binational science
foundation BSF \#1999332 and by the NSF grant DMR0097242.

\end{multicols}

\begin{thebibliography}{99}
\bibitem{ref goldman} A flattening of the R-T curve was reported in
reference 4. We do not observe this. We can create it artificially by
inducing noise to the film. The noise linearizes the non-linear part of the
I-V characteristic of the junction and dominates the temperature dependence
at low currents and voltages.
\end{thebibliography}
\end{document}